\documentclass[a4paper,fleqn]{cas-sc}

\usepackage[authoryear,longnamesfirst]{natbib}
\usepackage{graphicx}
\usepackage{booktabs}
\usepackage{multirow}
\usepackage{longtable, array}

\def\tsc#1{\csdef{#1}{\textsc{\lowercase{#1}}\xspace}}
\tsc{WGM}
\tsc{QE}

\begin{document}
\let\WriteBookmarks\relax
\def\floatpagepagefraction{1}
\def\textpagefraction{.001}

\setlength{\tabcolsep}{5pt}

\shorttitle{GLAT: The Generative AI Literacy Assessment Test}    

\shortauthors{Jin et al.}  

\title [mode = title]{GLAT: The Generative AI Literacy Assessment Test}  



%
\author[1]{Yueqiao Jin}[orcid=0009-0003-7309-4984]
\author[1]{Roberto Martinez-Maldonado}[orcid=0000-0002-8375-1816]
\author[1]{Dragan Gašević}[orcid=0000-0001-9265-1908]
\author[1]{Lixiang Yan}[orcid=0000-0003-3818-045X]
\cormark[1]







\affiliation[1]{organization={Monash University},
            addressline={25 Exhibition Walk}, 
            city={Clayton},
            postcode={3168}, 
            state={Victoria},
            country={Australia}}
            
\cortext[1]{Corresponding author}









\begin{abstract}
The rapid integration of generative artificial intelligence (GenAI) technology into education necessitates precise measurement of GenAI literacy to ensure that learners and educators possess the skills to engage with and critically evaluate this transformative technology effectively. Existing instruments often rely on self-reports, which may be biased. In this study, we present the GenAI Literacy Assessment Test (GLAT), a 20-item multiple-choice instrument developed following established procedures in psychological and educational measurement. Structural validity and reliability were confirmed with responses from 355 higher education students using classical test theory and item response theory, resulting in a reliable 2-parameter logistic (2PL) model (Cronbach's alpha = 0.80; omega total = 0.81) with a robust factor structure (RMSEA = 0.03; CFI = 0.97). Critically, GLAT scores were found to be significant predictors of learners' performance in GenAI-supported tasks, outperforming self-reported measures such as perceived ChatGPT proficiency and demonstrating external validity. These results suggest that GLAT offers a reliable and valid method for assessing GenAI literacy, with the potential to inform educational practices and policy decisions that aim to enhance learners' and educators' GenAI literacy, ultimately equipping them to navigate an AI-enhanced future.
\end{abstract}




\maketitle

\section{Introduction}

Generative artificial intelligence (GenAI) has rapidly emerged as a transformative force in higher education, challenging traditional pedagogical frameworks while simultaneously presenting novel opportunities for teaching, learning, and assessment. Tools like OpenAI's ChatGPT, Google's Gemini, and Anthropic's Claude hold the potential to transform how personalised tutoring service can be delivered, how instructional materials can be generated, how lectures can be transcribed for accessibility, and how creativity can be nurtured through multimedia content generation \citep{yan2024promises, khosravi2023generative}. However, the integration of these technologies is accompanied by complex challenges, including ethical considerations, risks of misinformation from model "hallucinations," and concerns regarding academic integrity \citep{ji2023survey, mcdonald2024generative}. Such complexities necessitate a deeper focus on fostering AI literacy, particularly GenAI literacy, among both educators and learners to fully harness GenAI's benefits and mitigate its associated risks \citep{ng2021conceptualizing, zhao2024chatgpt}.

AI literacy refers to the set of competencies that enable individuals to effectively interact with AI technologies, encompassing understanding fundamental AI concepts, engaging in critical evaluation, and using AI tools ethically in diverse contexts \citep{Long_2020, ng2021conceptualizing}. Within this broad framework, GenAI literacy emerges as a specialised subset, focusing on skills required to engage with GenAI systems that can autonomously produce text, visuals, or other forms of media \citep{yan2024promises, annapureddy2024generative}. Developing GenAI literacy involves more than just foundational knowledge; it requires proficiency in crafting prompts, interpreting AI-generated outputs, and understanding the socio-ethical implications of using such tools \citep{zhao2024chatgpt, Bozkurt_2024}. As GenAI becomes increasingly embedded in educational systems, it is imperative for learners and educators to acquire these competencies to effectively leverage the technology while minimising potential pitfalls such as biases or inaccuracies \citep{lyu2024evaluating, chiu2024future}.

Numerous instruments have been developed to assess AI literacy, reflecting the diversity of competencies that individuals need to navigate AI technologies. Conventional AI literacy assessments often rely on self-reported surveys, which are effective in capturing perceived knowledge but may lack the reliability needed to accurately measure actual competencies, especially given the tendency for individuals to overestimate their understanding \citep{Lintner_2024, laupichler2023delphi}. Most existing instruments address general AI literacy, focusing on technical knowledge, awareness, and ethical considerations, but fail to adequately capture the unique skills required for GenAI \citep{koch2024overview, zhao2024chatgpt}. There is a growing demand for more nuanced and context-specific instruments to evaluate GenAI literacy, particularly as generative tools become integral to both physical and digital learning environments \citep{koch2024overview, zhao2024chatgpt, yan2024practical}.

Current AI literacy assessments can be broadly categorised into two types: self-reported and performance-based measures. Self-reported instruments, while commonly used, provide insights into individuals' perceived abilities but may introduce biases that obscure a more reliable measure of literacy levels \citep{ng2021conceptualizing, Lintner_2024}. In contrast, performance-based assessments evaluate actual competencies through direct engagement, offering a more reliable measure of skills. This distinction is especially pertinent for GenAI literacy, where there is often a gap between learners' perceived understanding and their real ability to effectively utilise generative tools \citep{lyu2024evaluating}. GenAI technologies necessitate iterative, context-specific interactions that require both sophisticated prompting skills and the ability to critically assess AI outputs, areas where self-reports may fall short \citep{chiu2024future, Bozkurt_2024}. Therefore, developing performance-based instruments is essential to provide a reliable assessment of individuals' abilities to engage with these advanced technologies in educational settings. However, to the best of our knowledge, there are still limited performance-based tools for measuring students' GenAI literacy in higher education, particularly those that have been rigorously developed and validated according to established psychological and educational measurement standards \citep{thorndike1991measurement, aera2014standards}.

The current study contributes to the field of AI in education and AI literacy by introducing the GenAI Literacy Assessment Test (GLAT), a performance-based instrument specifically designed to evaluate GenAI literacy within higher education contexts. The GLAT aims to fill a critical gap in existing assessment tools by providing a more reliable, comprehensive evaluation of the key competencies required to interact with GenAI tools. Unlike existing assessments that focus predominantly on general AI skills, GLAT targets the unique skills necessary for effective engagement with generative technologies, including technical proficiency, ethical awareness, and the capacity for critical evaluation of GenAI-generated outputs. This instrument is grounded in rigorous methodologies from psychological and educational measurement \citep{thorndike1991measurement, aera2014standards}, ensuring both validity and reliability. By focusing on performance-based metrics, GLAT provides educators and researchers with a reliable tool to assess how well students and educators understand and can leverage GenAI technologies, ultimately informing targeted interventions that can enhance these competencies.
\section{Background}

\subsection{AI and GenAI Literacy}
The rapid advancements in AI technologies have accentuated the importance of AI literacy, especially in educational research. Researchers have endeavoured to define and conceptualise "AI literacy," with the definition by \citet[p.2]{Long_2020} being frequently cited: "a set of competencies that enables individuals to critically evaluate AI technologies; communicate and collaborate effectively with AI; and use AI as a tool online, at home, and in the workplace." Building upon this definition, other researchers have explored essential aspects of AI literacy. For instance, \citet{kandlhofer2016artificial} and \citet{burgsteiner2016irobot} concentrated on the comprehension of fundamental AI concepts present in various products and services. Meanwhile, \citet{wang2023measuring} emphasised critical evaluation, practical application, and ethical responsibilities. Additionally, \citet{ng2021conceptualizing}, \citet{ng2021ai}, and \citet{almatrafi2024systematic} have refined this framework by stressing competencies such as recognition, application, evaluation, creation, and ethical navigation.

The rise of GenAI technologies, like ChatGPT, necessitates a re-evaluation of AI literacy within the specific context of generative technologies. GenAI's capability to generate substantial content from minimal input alters the landscape, prompting a need for a revised understanding of AI literacy in this context \citep{zhao2024chatgpt}. Although there has been increasing academic interest in GenAI, a comprehensive definition of GenAI literacy remains elusive \citep{annapureddy2024generative}. Many current AI literacy frameworks are too general and do not address the specific competencies required by GenAI, which differ significantly from those of predictive models. Specifically, GenAI literacy calls for an integrative approach that combines theoretical knowledge, practical skills, and critical reflection. The 3wAI Framework by \citet{bozkurt2024generative} addresses this need, focusing on "Know What," "Know How," and "Know Why" and aiming to promote foundational understanding, practical application, and ethical awareness. \citet{zhao2024chatgpt} asserts that GenAI literacy should include pragmatic, safety, reflective, socio-ethical, and contextual elements. Scholars such as \citet{lyu2024evaluating} and \citet{annapureddy2024generative} emphasise the distinction between general AI literacy and the specific skills required for GenAI, pointing out that existing frameworks often overlook the skills necessary for effectively utilising these tools.

The need for GenAI literacy is particularly crucial in educational settings. The absence of a comprehensive GenAI literacy framework poses challenges to its effective integration into learning environments \citep{annapureddy2024generative}. \citet{chiu2024future} underlines the necessity of empirically evaluating pedagogies that incorporate GenAI to determine their impact on student outcomes. GenAI literacy can significantly enhance language learning, as suggested by \citet{alzubi2024generative}, while \citet{bozkurt2023unleashing} advocates for its inclusion in curricula to prepare students for an AI-augmented future. Despite the potential of GenAI tools to improve student learning, \citet{lyu2024evaluating} discovered that student-generated prompts often lack quality, highlighting a deficiency in necessary skills. This underscores the urgent need to develop GenAI literacy so that students can fully harness the potential of these technologies.

\subsection{AI Literacy Instruments}
Multiple AI literacy measurement instruments have been developed to address various contexts, audiences, and facets of AI literacy. These facets encompass technical, ethical, behavioural, and contextual elements, reflecting the multifaceted nature of AI literacy. The Scale for the Assessment of Non-Experts' AI Literacy (SNAIL), created by \citet{laupichler2023development}, evaluates technical knowledge, critical analysis, and practical application of AI. This scale's validity was confirmed through factor analyses and a Delphi study \citep{laupichler2023delphi}. The AI Literacy Scale (AILS) focuses on general AI literacy by measuring awareness, usage, evaluation, and ethics and was validated by subject matter experts \citep{wang2023measuring}. For specific audiences, the Medical Artificial Intelligence Readiness Scale for Medical Students (MAIRS-MS) targets medical students, assessing cognition, ability, vision, and ethics \citep{karaca2021medical}. Another survey, based on the Unified Theory of Acceptance and Use of Technology (UTAUT) and the Technological Pedagogical and Content Knowledge (TPACK) framework, examines pedagogical knowledge and AI use intentions among EFL teachers \citep{an2023modeling}. Instruments for younger audiences include the AI Literacy Questionnaire (AILQ) by Ng et al.\citeyear{ng2024design}, designed for secondary students, which assesses affective, behavioural, cognitive, and ethical dimensions. Chai et al.'s \citeyear{chai2021perceptions} AI Literacy Instrument explores students' confidence, readiness, and perceptions of AI. For broader competencies, Carolus et al.'s \citeyear{carolus2023mails} Meta AI Literacy Scale (MAILS) covers ethics, persuasion literacy, and emotion regulation. Additionally, Pinski et al.'s \citeyear{pinski2023ai} AI Literacy Instrument targets AI professionals, focusing on human-AI interaction, AI processing, and task knowledge, with validation for reliability and robustness. Lastly, Lee and Park's \citeyear{lee2024development} ChatGPT Literacy Tool is currently the only instrument specifically designed to assess GenAI skills among university students, though it relies on self-reported data.

Despite progress in AI literacy assessment, substantial gaps persist. Most current instruments rely heavily on self-reported assessments, with minimal use of performance-based measurements. A systematic literature review by \citet{Lintner_2024} identified 13 self-reported and only three performance-based instruments, highlighting the predominant reliance on self-reported tools and the urgent need for more reliable assessments. While there are some performance-based measures for general AI literacy, such as Hornsberger's \citeyear{hornberger2023university} test, which includes 30 multiple-choice questions and a sorting item, and Chiu's \citeyear{chiu2024developing} test consisting of 25 multiple-choice questions, no such measures have been developed for GenAI literacy. Performance-based evaluations are crucial for GenAI literacy, given the importance of practical engagement and iterative interactions \citep{Lintner_2024, laupichler2023delphi, yan2024promises}. To address this gap, it is essential to develop new GenAI literacy instruments specifically targeting GenAI skills and incorporating performance-based assessments for a more accurate evaluation of learners' competencies. These instruments would help educators better understand how students interact with generative models, identify areas needing additional training, and design effective interventions to enhance students' abilities to use GenAI technologies.

\subsection{Classical Test Theory and Item Response Theory}

Classical Test Theory (CTT) and Item Response Theory (IRT) offer complementary methodological approaches essential for developing robust and reliable assessment instruments in educational research \citep{de2010primer, thorndike1991measurement, hambleton1993comparison}. These theories provide the foundation for evaluating the structural validity and reliability of assessment tools, ensuring they accurately measure intended constructs such as GenAI literacy \citep{thorndike1991measurement, aera2014standards, de2010primer}. Specifically, CTT is grounded in the principle that an individual's observed test score is a combination of a true score and an error score. It provides a straightforward framework for analysing test data, focusing primarily on the reliability of test scores and the consistency of test items. Reliability in CTT is often assessed using measures such as Cronbach's alpha, which evaluates how well the items within a test measure the same construct \citep{miller1995coefficient}. CTT is instrumental in the initial stages of performance-based assessment development, ensuring that selected items reflect the latent constructs of interest, such as GenAI literacy \citep{de2010primer}. On the other hand, IRT offers a more sophisticated analysis by examining the relationship between an individual's ability and the probability of correctly responding to a test item. IRT provides detailed item-level information necessary for refining assessments, offering insights into how items function across varying levels of learner ability \citep{aera2014standards, de2010primer}. This approach allows for a comprehensive analysis of item characteristics such as difficulty and discrimination, ensuring that the test measures a wide range of skills effectively across diverse populations. Together, CTT and IRT provide a comprehensive framework that is crucial for the development and refinement of educational assessments. 

\subsection{External Validity and Domain Knowledge}

External validity is a crucial aspect of educational and psychological assessments, ensuring that the constructs measured by an instrument, such as GenAI literacy, can predict relevant learning outcomes and performances in varied contexts \citep{messick1995validity, boateng2018best}. This concept is emphasised in Messick's unified theory of validity, which posits that the validation process must consider not only how well an instrument measures the intended construct but also how well the construct aligns with real-world tasks and external criteria \citep{messick1995validity}. In the context of GenAI literacy, external validity involves the instrument's ability to accurately predict students' capacity to engage with and perform learning tasks using GenAI tools effectively. This predictive capability is critical, as the ultimate goal of assessing GenAI literacy is to ensure that learners are equipped with the skills necessary to apply these technologies in authentic educational and professional settings \citep{yan2024promises, annapureddy2024generative}. 

Domain knowledge plays a vital role in assessing the external validity of GenAI literacy tools \citep{alexander1992domain, alexander1988interaction}. It is essential to consider and control for the varying levels of domain knowledge possessed by students \citep{alexander1988interaction}, as this can significantly influence how effectively they can utilise GenAI technologies within different learning tasks \citep{tricot2014domain}. Failure to account for domain knowledge could result in skewed assessments \citep{alexander1992domain}, where the lack of subject-specific understanding might be mistaken for deficiencies in GenAI literacy itself. Consequently, assessing GenAI literacy necessitates a nuanced approach that considers the interplay between domain knowledge and the specific skills required to interact with GenAI tools. 

\subsection{Generative AI Literacy Assessment Test (GLAT)}
Building on these theoretical foundations, this study introduces the development and validation of the GLAT. Designed to measure GenAI literacy among higher education students, the GLAT stands on the robust foundations of psychological and educational measurement practices \citep{thorndike1991measurement, aera2014standards}. These foundations ensure that the GLAT can effectively assess key aspects of GenAI literacy, including foundational knowledge, application, ethical awareness, and critical evaluation capabilities \citep{annapureddy2024generative, Long_2020}. Specifically, the following research questions were investigated, aiming to assess the validity and reliability of the GLAT in capturing learners' GenAI literacy and predicting their learning performance:

\begin{itemize}
    \item RQ1: To what extent does the GLAT exhibit structural validity and reliability through classical test theory and item response theory?
    \item RQ2: To what extent does the GLAT measure of GenAI literacy demonstrate external validity by predicting learners' performance in learning tasks with GenAI chatbots compared to self-reported instruments?
\end{itemize}
\section{Methods}

The GLAT was developed following the established test development procedures outlined in Psychological and Educational Measurement \citep{thorndike1991measurement}. The development process involved: 1) creating a blueprint of relevant GenAI concepts, 2) generating an initial set of test items based on this blueprint, and 3) evaluating face and content validity through expert reviews and pilot studies, respectively. An item analysis was conducted using CTT, which involved selecting items based on item difficulty and discrimination index. The structural validity and reliability of the GLAT (RQ1) were assessed using IRT. The external validity of the GLAT (RQ2) was evaluated by analysing its effectiveness in predicting learners' performance on tasks involving interaction with a GenAI chatbot, compared to a self-reported GenAI literacy instrument \citep{lee2024development}. Details were elaborated in the following sections.

\subsection{Participants}
\begin{figure*}[htbp]
    \centering
    \includegraphics[width=0.8\linewidth]{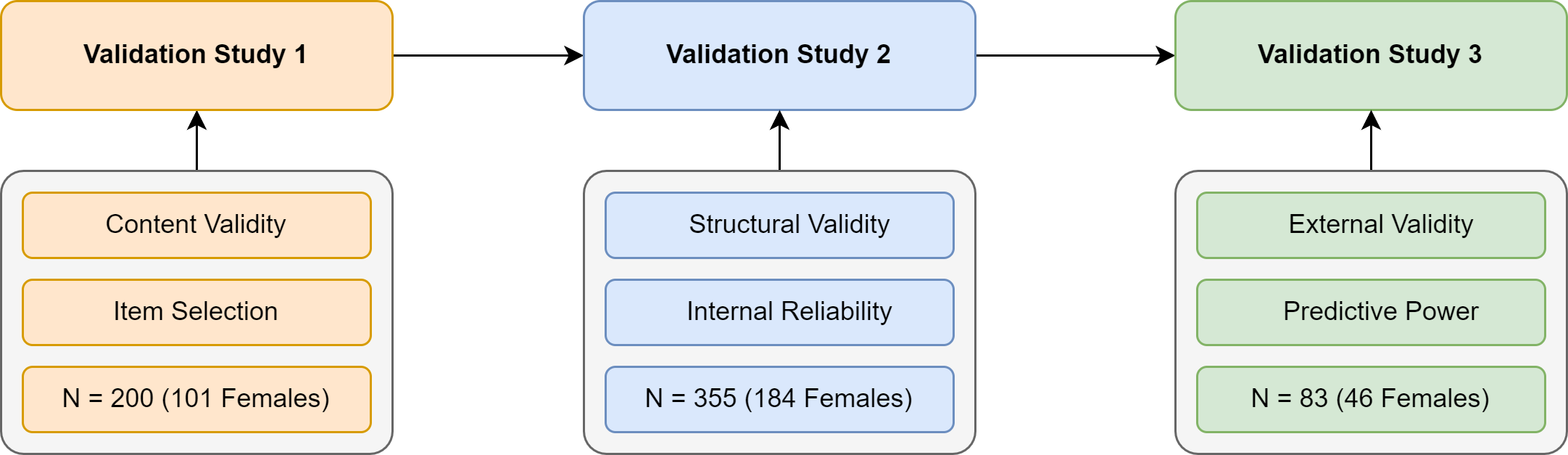}
    \caption{The participant sample size and focus of each validation study.}
    \label{fig-sample}
\end{figure*}

Three samples of higher education students were involved in various validation processes for the GLAT, following established standards \citep{thorndike1991measurement, aera2014standards}. As shown in Figure \ref{fig-sample}, the first validation study involved assessing the content validity and selecting the GLAT item using CTT. Responses were gathered from 200 higher education students (101 females). The second validation study aimed to evaluate the structural validity and reliability of the GLAT using IRT (RQ1). This study analysed responses from 355 higher education students (184 females). The final validation study focused on assessing the external validity of the GLAT (RQ2) and included 83 higher education students (46 females). All participants were recruited through Prolific\footnote{\url{https://www.prolific.co/}}, a reputable online research recruitment platform. Participants in the first and second studies received £1.5 for their time, while those in the final study were compensated £8 due to its increased complexity, as elaborated in Section \ref{rq2}. All studies were conducted using Qualtrics, with the item and option orders in the GLAT randomised \citep{aera2014standards}. Ethics approval was obtained from [Anonymised] University (Project ID: Anonymised), and informed consent was obtained from all participants.

\subsection{Item Generation}
\subsubsection{Blueprint construction}

The blueprint of the GLAT was developed based on the four dimensions of AI literacy proposed by \citet{ng2021conceptualizing}, including 1) Know \& Understand, 2) Use \& Apply, 3) Evaluate \& Create, and 4) Ethics, and focusing on the specific context of GenAI. To identify a set of relevant GenAI concepts in each dimension, we cover a wide range of resources, including academic publications in prestige journals (e.g., Nature and Science), reports and articles published by reputable organisations (e.g., UNESCO, MIT News, and Standford HAI), and education information released by leading GenAI technology companies (e.g., OpenAI, Google, Meta, and NVIDIA). The decision to include diverse sources, beyond traditional academic publications, was driven by the rapidly evolving landscape of GenAI. This encompasses foundational models (e.g., large language models and diffusion models) as well as supporting infrastructure and techniques (e.g., embedding databases and retrieval methods). Our process for extracting relevant concepts involved three steps. Initially, two researchers independently reviewed the source documents (n=19; links are available in the repository) and recorded pertinent GenAI concepts. Subsequently, they collaborated to consolidate similar concepts, resulting in a refined set of 25 concepts (see Table \ref{tab-concepts}). A validation panel of three GenAI researchers then reviewed these concepts to ensure they provided reasonable coverage of the latest developments in GenAI.

\begin{table}[htbp]
\centering
\renewcommand{\arraystretch}{1.2}
\caption{Dimensions and concepts related to generative AI.}
\begin{tabular}{|p{2cm}|p{13.5cm}|}
\hline
\textbf{Dimension} & \textbf{Concepts} \\ \hline
Know \& \newline Understand & Generative AI (GenAI), Foundation model (e.g., LLM and diffusion model), Generation capability, Zero-shot learning, Prompt-based development, Content generation, Token, Artificial General Intelligence (AGI), Model alignment, RAG (Retrieval-Augmented Generation) \\ \hline
Use \& \newline Apply & Contextual understanding in content creation, Knowledge updating and integration, Information retrieval and synthesis, Token management and limitation, Multimodal content generation \\ \hline
Evaluate \& \newline Create & Trustworthiness of LLM outputs, LLM knowledge cutoff, Information cross-check (Hallucination), GenAI-generated content authenticity, Voice cloning with GenAI \\ \hline
Ethics & Model biases, Black box issue in GenAI, Copyright issues in GenAI-generated content, Content safety, Privacy concerns with GenAI \\ \hline
\end{tabular}
\label{tab-concepts}
\end{table}

\subsubsection{Item generation}

The item generation process began with aligning each item to the specific GenAI literacy concepts outlined in the blueprint (Table \ref{tab-concepts}). This ensured comprehensive coverage across all dimensions of GenAI literacy, including knowledge, application, evaluation, and ethics. Each item was crafted as a multiple-choice question (MCQ) to assess understanding through a consistent and structured format, following the established guidelines \citep{haladyna2004developing}. MCQs are particularly effective for evaluating knowledge across large participant groups due to their standardised nature and ease of scoring. A critical component of MCQ design is the creation of plausible distractors -- incorrect answer options that are designed to challenge and differentiate between varying levels of participant understanding \citep{haladyna2002review}. Each distractor was carefully developed to reflect common misconceptions or logical errors relevant to the GenAI concept being assessed, ensuring they were plausible enough to create meaningful distinctions in responses. The drafting process involved multiple revisions to enhance the clarity, relevance, and cognitive demand of each question. Two researchers iteratively refined and checked the questions for ambiguity and ensured the options were free of overlapping meanings or unintended cues. An expert panel comprising specialists in GenAI, educational psychology, and psychometrics also reviewed each item and provided feedback for improvements. Eventually, the initial 25-item version of the GLAT was developed and assessed for content validity (Table \ref{tab-glat}).

\begin{footnotesize}
\renewcommand{\arraystretch}{1.2}
\begin{longtable}{|p{0.5cm}|p{1.2cm}|p{5.0cm}|p{8.2cm}|}
\caption{The initial 25-item version of the Generative AI Literacy Assessment Test (GLAT).} 
\label{tab-glat}\\
\hline
\textbf{Item} & \textbf{Dimension} & \textbf{Question} & \textbf{Options (Answer Highlighted)} \\
\hline
1 & Know \& \newline Understand & Which of the following best describes "Generative AI"? & 
\textbf{A. AI that creates new content like text, images, or music by learning from existing data.} \newline
B. An AI system designed to enhance the speed and accuracy of data retrieval in search engines. \newline
C. A form of artificial intelligence that focuses on translating languages in real-time. \newline
D. AI technology used primarily for managing and organizing large databases. \\
\hline

2 & Know \& \newline Understand & Which of the following statements best describes an LLM (Large Language Model)? & 
A. It generates text by analyzing and summarizing large volumes of web content. \newline
\textbf{B. It generates text by predicting the next word based on the context of previous words.} \newline
C. It generates text by translating input text into multiple languages simultaneously. \newline
D. It generates text by using pre-defined templates and filling in the blanks. \\
\hline

3 & Know \& \newline Understand & Which of the following tasks can Generative AI perform with a high degree of accuracy? & 
A. Predicting stock market trends \newline
B. Making ethical decisions in complex scenarios \newline
C. Diagnosing rare diseases \newline
\textbf{D. Generating human-like text based on prompts} \\
\hline

4 & Know \& \newline Understand & In the context of Generative AI, what is "zero-shot learning"? & 
A. Training a model without any data. \newline
\textbf{B. The ability of a model to perform a task without any task-specific training.} \newline
C. A method of reducing the model's training time to zero. \newline
D. A technique for generating synthetic training data. \\
\hline

5 & Know \& \newline Understand & Which of the following is a potential challenge when using prompt-based development for text generation? & 
A. The language model can only generate binary outputs. \newline
B. The need for extensive labeled data to train the model. \newline
\textbf{C. Crafting a prompt that accurately captures the desired context and nuances.} \newline
D. The requirement for complex feature engineering. \\
\hline

6 & Know \& \newline Understand & \textcolor{red}{\textbf{[DROPPED]}} When using a generative AI model to classify text into multiple categories, what is a common approach to handle more than two output classes? & 
A. Use multiple binary classifiers for each category. \newline
\textbf{B. Use a single prompt that includes all possible categories.} \newline
C. Train a separate model for each category. \newline
D. Use unsupervised learning to cluster the text data. \\
\hline

7 & Know \& \newline Understand & What does the term "token" refer to in the context of a large language model (LLM)? & 
\textbf{A. A token is a unit of text, such as a word or a subword, that the model processes individually.} \newline
B. A token is a unique identifier assigned to each user interacting with the language model. \newline
C. A token is a security measure used to authenticate API requests to the language model. \newline
D. A token is a reward given to users for contributing valuable data to train the language model. \\
\hline

8 & Know \& \newline Understand & Which of the following is NOT a requirement for an AI to be considered artificial general intelligence (AGI)? & 
A. The ability to learn and adapt to new tasks without human intervention. \newline
B. The capability to perform tasks across various domains with human-like proficiency. \newline
\textbf{C. The ability to predict future events with perfect accuracy.} \newline
D. The capacity to understand and generate natural language. \\
\hline

9 & Know \& \newline Understand & \textcolor{red}{\textbf{[DROPPED]}} Why is model alignment important in the development of generative AI? & 
\textbf{A. To ensure AI systems better reflect human values and are safer.} \newline
B. To improve computational efficiency and reduce energy consumption. \newline
C. To enhance the alignment between model responses and user requests. \newline
D. To increase the speed of data processing and analysis. \\
\hline

10 & Know \& \newline Understand & How does RAG (Retrieval-Augmented Generation) enhance the capabilities of an LLM? & 
A. By improving its grammar and syntax. \newline
\textbf{B. By providing it with real-time and relevant data.} \newline
C. By increasing its computational speed. \newline
D. By enabling it to understand multiple languages. \\
\hline

11 & Use \& \newline Apply & When using generative AI to create a marketing pitch, which of the following strategies is least likely to be effective? & 
A. Supplying the AI with information about the target audience \newline
B. Asking the AI to include unique selling points and benefits \newline
C. Requesting the AI to use persuasive language techniques \newline
\textbf{D. Providing the AI with a list of competitors' products} \\
\hline

12 & Use \& \newline Apply & After deploying a customer service chatbot, you notice that it frequently provides outdated information about company policies. What is the best course of action to address this issue? & 
A. Implement a feedback loop where users can flag outdated information for review. \newline
\textbf{B. Schedule regular updates to the chatbot's training data to include the latest company policies.} \newline
C. Set up a system where complex or policy-related queries are escalated to human agents for accurate responses. \newline
D. Conduct a comprehensive audit of the chatbot's performance metrics to identify areas for improvement. \\
\hline

13 & Use \& \newline Apply & Suppose you have a large dataset of emails and you want to build an application to answer questions based on this dataset. Which of the following scenarios best illustrates the advantage of using RAG over prompting (i.e., without RAG)? & 
A. You need to generate creative writing pieces based on the email content. \newline
B. You want to ensure the model can answer questions even if it has never seen similar questions before. \newline
\textbf{C. You need to answer questions that require specific information from different parts of the email dataset.} \newline
D. You want to reduce the size of the language model to save computational resources. \\
\hline

14 & Use \& \newline Apply & \textcolor{red}{\textbf{[DROPPED]}} While using a Generative AI tool to write a story, you notice that the context window is limited to 500 tokens. What is a potential consequence of exceeding this limit? & 
A. The AI will automatically expand the context window \newline
B. The AI will ignore the excess tokens and generate text based on the first 500 tokens \newline
\textbf{C. The AI will generate text based on the most recent 500 tokens} \newline
D. The AI will stop functioning until the context window is reduced \\
\hline

15 & Use \& \newline Apply & \textcolor{red}{\textbf{[DROPPED]}} When creating a video with a generative AI tool that supports text, images, and audio narration, which feature is most critical for ensuring the tool can handle this task effectively? & 
\textbf{A. Text-to-Speech (TTS) capability.} \newline
B. Image recognition capability. \newline
C. Multilingual support. \newline
D. Sentiment analysis capability. \\
\hline

16 & Evaluate \& \newline Create & As a student using a Large Language Model (LLM) to gather information for an assignment, how should you approach the information it provides? & 
A. The LLM's answers are always more trustworthy than any information you will find on the internet, so you can use them without further verification. \newline
B. The LLM's answers are generally more trustworthy than internet sources, but you should still verify the information with other reliable sources. \newline
\textbf{C. The LLM's answers are not necessarily more trustworthy than internet sources, and you should cross-check the information with other credible references.} \newline
D. The LLM's answers are less trustworthy than internet sources because it relies on outdated information. \\
\hline

17 & Evaluate \& \newline Create & It is unlikely for an LLM to provide an accurate summary of the latest financial market trends in real-time. Is this statement true or false? & 
\textbf{A. True, because the LLM's data may be outdated due to its knowledge cutoff.} \newline
B. True, because the LLM is not good at handling numbers and structured data. \newline
C. False, because the LLM frequently updates its knowledge base. \newline
D. False, because the LLM is capable of synthesizing the latest market data automatically. \\
\hline

18 & Evaluate \& \newline Create & A generative AI tool has provided a summary of a research paper. The summary states, "The study found that increased screen time is directly correlated with decreased attention spans in children aged 8-12." What is your next step? & 
A. Accept the summary as accurate because AI tools are generally reliable. \newline
B. Ask the AI to provide more details about the study's methodology and results. \newline
\textbf{C. Cross-check the summary with the original research paper.} \newline
D. Use another AI tool to generate a summary for comparison and evaluate the consistency between both summaries \\
\hline

19 & Evaluate \& \newline Create & While reviewing a video of a well-known public figure making controversial statements, which characteristic confirms the video was NOT generated by AI? & 
A. The public figure's voice sounds like themselves. \newline
B. The video has a professional and polished appearance. \newline
C. The video is high-quality with smooth transitions. \newline
\textbf{D. None of the above.} \\
\hline

20 & Evaluate \& \newline Create & \textcolor{red}{\textbf{[DROPPED]}} Imagine someone is trying to clone your voice using generative AI. Which of the following audio recordings would be most effective? & 
A. A 1-hour recording of your conversation in a noisy public space. \newline
\textbf{B. A 30-second recording of your presentation speech.} \newline
C. A 5-minute recording of you singing a pop song. \newline
D. Generative AI cannot be used to clone human voice. \\
\hline

21 & Ethics & When a generative AI system is used for screening job applications, what issue might arise concerning the quality and fairness of hiring decisions? & 
A. The AI system might overlook applicants' unique achievements and extracurricular activities. \newline
B. The AI system could misinterpret minor formatting differences in resumes. \newline
C. The AI system might not effectively handle applications submitted in various languages. \newline
\textbf{D. The AI system could reinforce existing biases found in historical hiring data.} \\
\hline

22 & Ethics & In a healthcare startup, an accurate AI model recommends treatments, but doctors don't trust it because they can't understand how the model arrived at its conclusions. What core issue does this scenario illustrate? & 
A. The AI model uses obsolete training data. \newline
B. The training dataset lacks sufficient diversity. \newline
C. The treatment guidelines input are incorrect. \newline
\textbf{D. The AI model behaves as a black box.} \\
\hline

23 & Ethics & What are the potential copyright implications for a journalist using an AI-generated image in a commercial article? & 
\textbf{A. The journalist needs to check the licensing policy of the AI tool they used.} \newline
B. The AI-generated image is automatically free to use without any restrictions. \newline
C. The journalist must pay a standard licensing fee to use the AI-generated image. \newline
D. The image cannot be used in any commercial context because it is AI-generated. \\
\hline

24 & Ethics & Should we impose restrictions on the outputs of generative AI technologies? & 
A. Yes, to reduce the computational resources required for operating these technologies. \newline
\textbf{B. Yes, to prevent the dissemination of harmful or misleading content.} \newline
C. No, as it would hinder technological innovation and creativity. \newline
D. No, because users should have the freedom to access all generated content. \\
\hline

25 & Ethics & Sending personal information to cloud-based generative AI tools has little privacy concerns. & 
A. True, as this information is encrypted using sophisticated algorithms during the transmission process. \newline
B. True, as generative AI tools are black-box systems and cannot output personal information even if it is used for model training. \newline
\textbf{C. False, as generative AI tools train on unencrypted data and can output private information based on their probabilistic nature.} \newline
D. False, as advancements in quantum computing can easily decipher the encrypted data. \\
\hline
\end{longtable}
\end{footnotesize}

\subsubsection{Content validity.}

The content validity of the GLAT was assessed in a pilot study with a sample of 200 higher education students through six questions (Table \ref{tab-content}) based on the Standards for Educational and Psychological Testing \citep{aera2014standards}. These questions assessed content validity from multiple aspects (Cronbach's alpha = 0.81), including relevance (C1), comprehensiveness (C2 and C3), comprehensibility (C4 and C5), and face validity (C6). Each question was measured using a five-point Likert scale, ranging from \textit{strongly disagree} (1) to \textit{strongly agree} (5). As shown in Table \ref{tab-content}, the GLAT demonstrated strong content validity in all four aspects based on the pilot study results. This indicates that the GLAT is a valid tool for assessing GenAI literacy, as it effectively covers the necessary content areas, is easy to understand, and is perceived as an effective assessment tool by the students. This strong content validity sets the stage for the next phase of test development: ensuring structural validity and reliability through rigorous item selection.

\begin{table}[htbp]
\centering
\caption{Content validity questions for the Generative AI Literacy Assessment Test (GLAT).}
\label{tab-content}
\renewcommand{\arraystretch}{1.2}
\begin{tabular}{|m{1.5cm}|m{10cm}|m{1.5cm}|m{1.5cm}|}
\hline
\textbf{Question} & \textbf{Detail} & \textbf{Mean} & \textbf{SD} \\
\hline
C1 & The questions are directly related to generative AI concepts and skills. & 4.62 & 0.65 \\
\hline
C2 & The test covers a broad range of concepts that are necessary for assessing generative AI literacy. & 4.35 & 0.72 \\
\hline
C3 & The test includes questions that assess both fundamental and advanced concepts. & 4.46 & 0.74 \\
\hline
C4 & The questions are clearly written and easy to understand. & 4.26 & 0.79 \\
\hline
C5 & The options provided for each question are clearly distinct and easily distinguishable. & 4.28 & 0.85 \\
\hline
C6 & Overall, I believe the test is an effective tool for assessing generative AI literacy. & 4.22 & 0.87 \\
\hline
\end{tabular}
\end{table}

\subsection{RQ1: Structural Validity and Reliability}
\subsubsection{Item selection}
The item selection process utilised CTT, focusing on two key metrics: item difficulty and the discrimination index \citep{hambleton1993comparison}. Item difficulty was determined as the proportion of participants who answered an item correctly, with values ranging from 0 to 1. The discrimination index, quantified by the point-biserial correlation, reflects an item's ability to differentiate between high and low performers on the test. To calculate the discrimination index, we subtracted the number of test-takers in the lower group who answered the item correctly from the number of test-takers in the upper group who did so, then divided the result by the total number of test-takers. This index ranges from -1 to 1. As the GLAT aims to provide a continuous measure of GenAI literacy across various proficiency levels, we reported item difficulty without establishing a specific criterion \citep{de2010primer}. However, items with a discrimination index below 0.3 were excluded to ensure that the final set of items effectively differentiated among test-takers \citep{oosterhof2001classroom}.

\subsubsection{Structural Validity}
\textbf{IRT models.} After eliminating items with low discrimination indices, we assessed the structural validity of the final item set based on IRT \citep{reise2009item}. Three different IRT models were used: the Rasch model, the 2-parameter logistic (2PL) model, and the 3-parameter logistic (3PL) model. These models provide a nuanced understanding of the relationship between item characteristics and the latent trait being measured, allowing for an examination of each item's difficulty (b-parameter), discrimination (a-parameter), and guessing (c-parameter). Specifically, the Rasch model assumes that all items have the same discrimination and that guessing is not a factor, focusing solely on item difficulty. The 2PL model extends this by allowing each item to have its own discrimination parameter, which can help capture variations in how well different items differentiate between test-takers of different ability levels. The 3PL model further includes a guessing parameter (25\% for four options MCQs), acknowledging that respondents may have a chance of answering an item correctly by guessing, especially in multiple-choice formats. The most appropriate IRT model was selected based on a comprehensive evaluation of model and item fit indexes, further elaborated below.

\textbf{Assumption test.} Before fitting IRT the models and evaluating model fit, a confirmatory factor analysis (single-factor model) was performed using the \textit{lavaan} package in R \citep{rosseel2012lavaan} to assess the assumption of unidimensionality \citep{reise2009item}. This analysis verifies whether a single latent construct, GenAI literacy, could adequately explain all items in the GLAT. The model was considered unidimensional if it met the following criteria: \(\chi^2\)/\textit{df} < 2, root mean square error of approximation (RMSEA) < 0.05, and standardized root mean square residual (SRMSR) < 0.1 \citep{brown2015confirmatory}. In addition, the assumption of local independence was evaluated using the Q3 statistic \citep{yen1984effects}. A threshold value of 0.2 was applied \citep{chen1997local}, with any pair of residual correlations exceeding this level indicating a potential violation of local independence.

\textbf{Model fit.} To evaluate model fit, the three IRT models were fitted using the \textit{mirt} package in R \citep{chalmers2012mirt}. we utilised several fit statistics commonly used in comparing IRT models \citep{reise2009item}. The primary statistic used for model comparison was the likelihood ratio test, which was conducted using an analysis of variance (ANOVA) framework to compare nested models. Specifically, we compared the simpler Rasch model to the more complex 2PL model, and then the 2PL model to the 3PL model. The ANOVA test provides insight into whether the increased complexity of a model significantly improves the fit of the data. This involves comparing the deviance (twice the negative log-likelihood) of each model, with a significant p-value indicating that the extra parameters provide a better fit. Alongside the likelihood ratio tests, we also assessed comparative fit using information criteria, specifically, the Akaike Information Criterion (AIC) and the Bayesian Information Criterion (BIC). These criteria take into account both the goodness of fit and the complexity of the model, with lower values suggesting a more preferable model. Additionally, we evaluated overall model fit with the M2 Statistic, RMSEA with values less than 0.06 indicating good fit, SRMSR with values below 0.08 suggesting good fit, and Tucker-Lewis Index (TLI) and Comparative Fit Index (CFI) with values above 0.90 indicating acceptable fit and above 0.95 suggesting excellent fit \citep{reise2009item, maydeu2013goodness}. 

\textbf{Item characteristic curves (ICCs).} ICCs were plotted for each model to assess each item's ability to capture the latent trait of GenAI literacy consistently across varying proficiency levels \citep{reise2009item}.

\textbf{Item fit.} The signed chi-squared (\(S\text{-}\chi^2\)) statistic was used to evaluate the item fit for each model \citep{orlando2000likelihood}. This method assesses how well the data fit the expected model by examining the extent of the difference between observed and expected response patterns for each item. To account for multiple tests and control the false discovery rate, the Benjamini-Hochberg procedure was applied to adjust the p-values \citep{benjamini1995controlling}.

\subsubsection{Reliability evaluation}
The reliability of the final item set was evaluated by focusing on internal consistency using Cronbach's alpha. Cronbach's alpha provides an estimate of the proportion of variance in the test scores attributable to the true score variance. A Cronbach’s alpha value of 0.7 or higher was deemed indicative of acceptable reliability in educational and psychological testing contexts, suggesting that the test items consistently assess the underlying construct of GenAI literacy \citep{miller1995coefficient}. In addition, we utilised the coefficient omega, which is considered a more recent and potentially more accurate measure of internal consistency, particularly when items have varying loadings on the construct \citep{dunn2014alpha}. Specifically, we used omega total to evaluate the overall reliability of the GLAT in measuring GenAI literacy. This coefficient, like Cronbach's alpha, ranges from 0 to 1 and utilises a similar threshold value (e.g., 0.7) to indicate acceptable reliability. Additionally, to further understand and confirm the precision and reliability of the GLAT, we evaluated the test information function \citep{reise2009item}. The test information function was analysed to ensure consistent measurement at proficiency levels most relevant to the GLAT’s intended application, particularly for learners and educators with low to average levels of GenAI literacy, considering GenAI technology is relatively new for them and in higher education \citep{annapureddy2024generative, jin2024generative}.

\subsection{RQ2: External Validity}
\label{rq2}
The external validity of the GLAT was assessed by analysing its predictive power concerning learners' task performance during a task that involved interacting with a GenAI-powered conversational chatbot. The choice of a task involving interaction with a GenAI-powered chatbot is particularly appropriate for assessing external validity, as it closely mirrors real-world applications and challenges students might encounter in educational settings \citep{mcgrath2024generative}, thereby providing a practical context for evaluating the external validity of GenAI literacy in predicting learner performance. This predictive capability was compared to that of a validated self-report instrument, the ChatGPT Literacy Scale \citep{lee2024development}, to determine whether the GLAT provides additional predictive value beyond the self-report measure. The ChatGPT literacy scale serves as a suitable comparator, as it is specifically designed and validated for higher education students in the context of interactions with ChatGPT, a GenAI-powered conversational chatbot. Further details regarding the learning task, study procedure, and the analytical methods used are provided below.

\subsubsection{Learning task}
\begin{figure*}[htbp]
    \centering
    \includegraphics[width=0.99\linewidth]{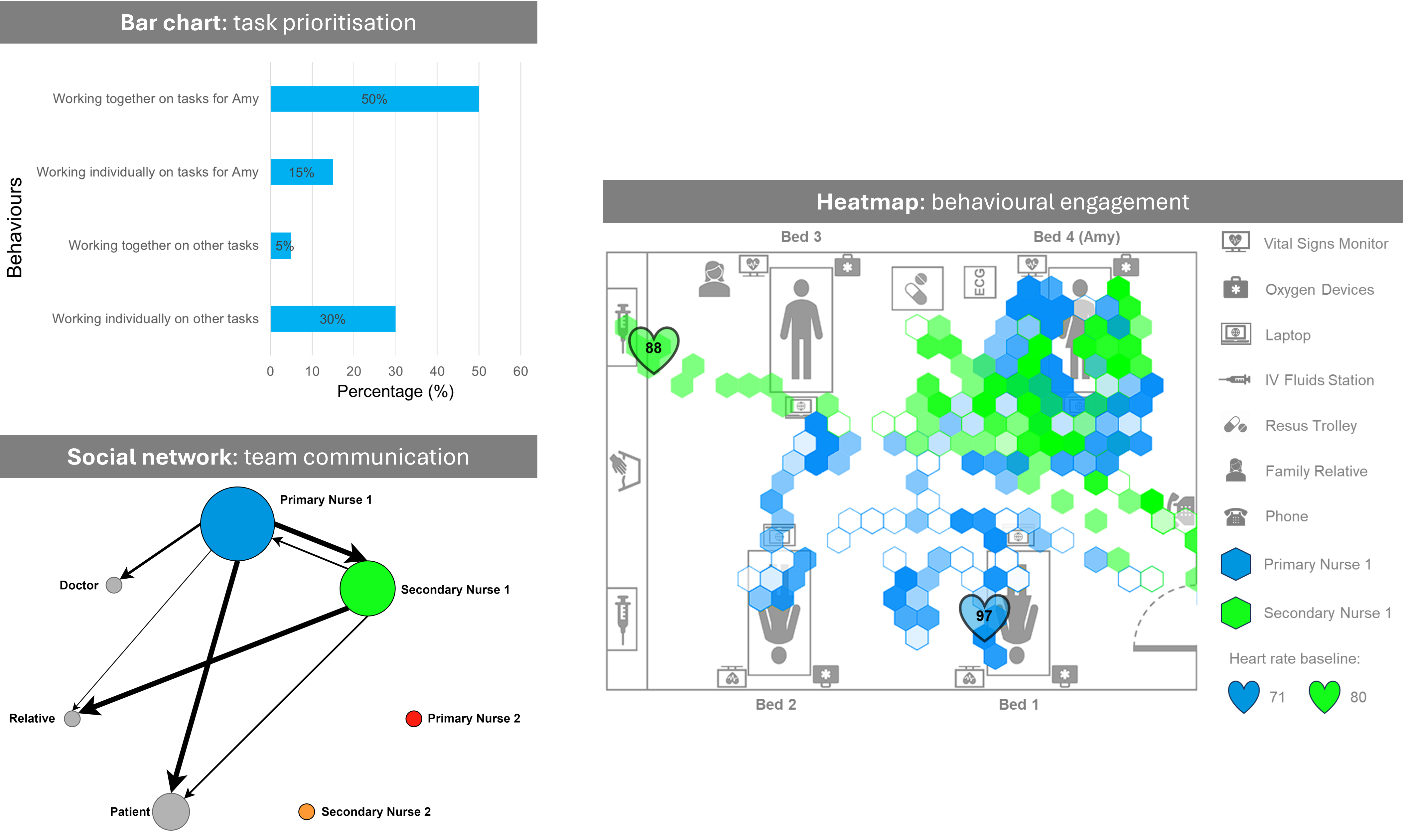}
    \caption{Visual analytics on teamwork in healthcare simulations, including: a) a bar chart of four prioritisation strategies, b) a social network diagram of communication behaviours among the actors, and c) a ward map showing individuals' physical positions (hexagon), verbal communication duration (colour saturation), and peak heart rate locations.}
    \label{fig-viz}
\end{figure*}

Learners engaged in a task that aimed to enhance their ability to comprehend complex visual analytics, an ability that many are lacking \citep{maltese2015data, Donohoe2020}, through interacting with GenAI chatbots. The learning task involved learners analysing a set of visual analytics on students' teamwork in healthcare simulations, composing a 100-word response and answering six evaluation questions to assess their ability to comprehend complex visual data. The visual analytics included three types of visualisations: a bar chart, a social network diagram, and a heatmap (Figure \ref{fig-viz}). Specifically, the bar chart illustrated students' prioritisation strategies with positional data, simplifying the comparison of time spent on behaviours during the simulation [Anonymised]. The social network (sociogram) mapped interaction patterns via positional and audio data, highlighting communication frequencies and directions with the patient, doctor, and relative [Anonymised]. The advanced heatmap map combined students' physical positions, verbal durations, and peak heart rate locations. Inspired by sports analytics \citep{goldsberry2012courtvision}, it used heatmaps to show verbal communication frequency and spatial distribution, and identified areas of peak physiological arousal.

A total of 83 higher education students (46 females) with medical, healthcare, and nursing backgrounds were involved in this validation to control for their familiarity with the healthcare simulation context. Learners were instructed to first analyse the visualisations and write a 100-word response on how the two nurses managed the primary patient, Amy, while attending to other beds, focusing on task prioritisation, verbal communication, and stress levels. After this, they answered six multiple-choice questions (two per visualisation) designed to assess comprehension of the visual data, addressing the first two levels of Bloom's taxonomy (knowledge and comprehension) \citep{bloom1984bloom}. For the knowledge questions, participants identified specific data points or patterns in visualisations, like determining which prioritisation behaviour two nurses spent the most time on from a bar chart. These questions assessed information retrieval skills. The comprehension questions required participants to interpret and derive insights, such as comparing spatial and verbal activities between two nurses using a ward map. These questions evaluated the ability to interpret insights and identify inconsistencies (see Table \ref{tab-question} for examples). Higher levels of Bloom's taxonomy, like application, were not considered because the learners have limited contextual knowledge of visual analytics.

\begin{table*}[htbp]
\centering
\caption{Example knowledge and comprehension question for the bar chart.}
{
\renewcommand{\arraystretch}{1.2}
\begin{tabular}{|l|l|}
\hline
\textbf{Bloom's Level} & \textbf{Question} \\ \hline
Knowledge & Which behaviour did the two nurses spend the \textit{most} time on? \\ \hline
Comprehension & How did the nurses spend their time working on tasks for Amy compared to other tasks? \\ \hline
\end{tabular}
}
\label{tab-question}
\end{table*}

\subsubsection{Chatbot design}
\begin{figure*}[htbp]
    \centering
    \includegraphics[width=0.5\linewidth]{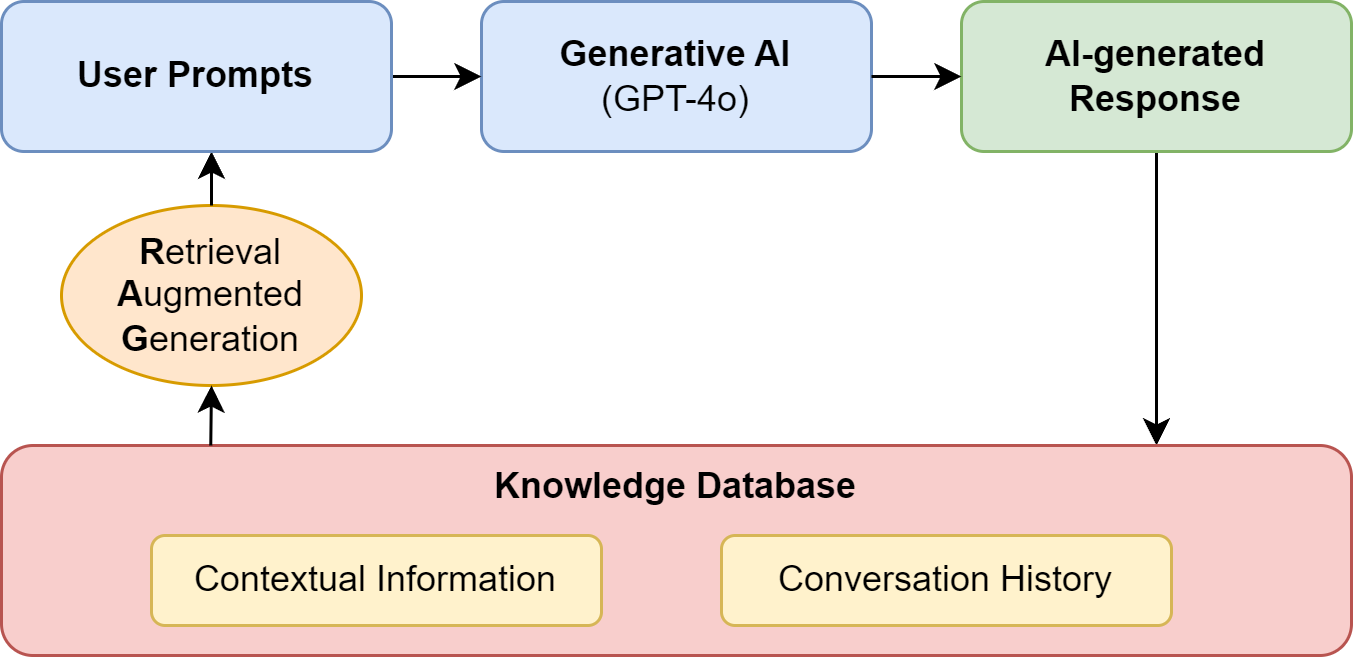}
    \caption{System design of the generative AI (GenAI) chatbots.}
    \label{fig-chatbot}
\end{figure*}

The GenAI chatbot was designed using the state-of-the-art retrieval-augmented generation (RAG) approach to improve response relevance and reduce inaccuracies (hallucinations) \citep{siriwardhana2023improving}. As illustrated in Figure \ref{fig-chatbot}, when learners ask a question about the visual analytics, the chatbot first retrieves relevant contextual information. It does this by computing vector embeddings of the prompts and calculating the cosine similarity between these prompt embeddings and stored knowledge embeddings \citep{li2024matching}. The retrieved information, along with learners' questions and chat history, is then sent to a generative AI, specifically GPT-4o, to generate responses. The conversation is subsequently stored in the knowledge base to provide context for future interactions. This chatbot design is a representation of openly accessible GenAI chatbots, such as ChatGPT, Gemini, and Claude \citep{achiam2023gpt}.

\subsubsection{Study procedure and measures}
\begin{figure*}[htbp]
    \centering
    \includegraphics[width=0.8\linewidth]{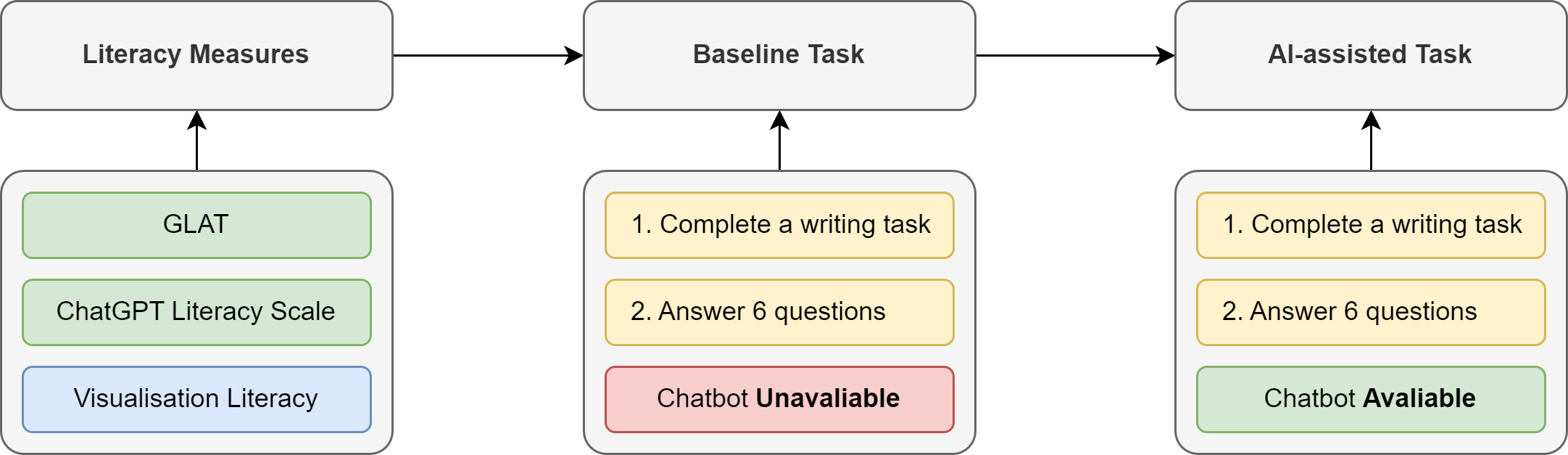}
    \caption{Study design: three literacy measurements, a baseline task, and an AI-assisted task.}
    \label{fig-study}
\end{figure*}

As illustrated in Figure \ref{fig-study}, learners begin by completing three literacy assessments: 1) the GLAT, 2) the ChatGPT Literacy Scale \citep{lee2024development}, and 3) the mini-VLAT \citep{pandey2023mini}, which evaluates domain knowledge, specifically, learners' visualisation literacy pertinent to the task. Following a within-subject design, learners first perform the learning task independently, serving as the baseline condition. Subsequently, they repeat the task using different visual analytics (identical visualisations but with different data) while receiving support from the GenAI chatbot, constituting the AI-assisted condition. Consequently, five measures were captured and analysed: performance-based GenAI literacy using the GLAT (\textit{GLAT-literacy}), self-reported GenAI literacy via the ChatGPT Literacy Scale (\textit{ChatGPT-literacy}), visualisation literacy through the mini-VLAT (\textit{VLAT-literacy}), baseline task performance without support (\textit{baseline score}), and task performance with GenAI chatbot assistance (\textit{AI-assisted score}).

\subsubsection{Predictive analysis}
Predictive modelling was conducted to evaluate the external validity of the GLAT in predicting task performance compared to the ChatGPT Literacy Scale. Specifically, we used ordinary least squares (OLS) regression to examine the predictive power of learners' \textit{GLAT-literacy} and \textit{ChatGPT-literacy} (independent variables; IVs) on their \textit{AI-assisted score} (dependent variable; DV), while controlling for their \textit{baseline score} and \textit{VLAT-literacy} (independent variables; IVs). Each measure was first standardised to ensure a uniform format for interpretation. Interaction terms were excluded as an ANOVA revealed no significant model improvements (\( F(11, 67) = 1.45, p = .17 \)). The final regression included an intercept (\(\beta_0\)) and a main effect for each IV (\(\beta_1\) to \(\beta_4\)). All assumptions for the regression analysis were verified. Linearity was confirmed by plotting predicted versus observed values. The normality of residuals was checked using the Shapiro-Wilk test and QQ plots. Homoscedasticity was assessed with the Breusch-Pagan test, and the independence of residuals was evaluated using the Durbin-Watson test. All assumptions were satisfied. 

\begin{small}
\begin{equation}
\begin{aligned}
\text{\textit{AI-assisted score}} &= \beta_0 + \beta_1 \times \text{\textit{GLAT-literacy}} + \beta_2 \times \text{\textit{ChatGPT-literacy}} + \beta_3 \times \text{\textit{baseline score}} + \beta_4 \times \text{\textit{VLAT-literacy}}
\end{aligned}
\end{equation}
\end{small}
\section{Results}

\subsection{Structural Validity and Reliability (RQ1)}
\subsubsection{Item selection}
The item selection process for the GenAI Literacy Assessment Test (GLAT) revealed varying degrees of item difficulty and discrimination indices that facilitated the identification of items for inclusion in the final assessment. As shown in Table \ref{table-item_selection}, a total of five items (Items 6, 9, 14, 15, and 20) had discrimination indices below the threshold of 0.3. These items were consequently excluded from the final item set to ensure that the assessment effectively differentiates among test-takers. Specifically, Item 6 had a discrimination index of 0.06, Item 9 had a discrimination index of 0.25, Item 14 had a discrimination index of 0.27, Item 15 had a discrimination index of 0.21, and Item 20 had a discrimination index of 0.23. The remaining items demonstrated adequate discriminability, with indices ranging from 0.33 to 0.55 (M = 0.41, SD = 0.06), indicating that the retained items have a consistent ability to differentiate between high and low performers, with relatively low variability in their discriminative power. Of the 20 items retained, the item difficulties ranged between 0.25 and 0.90 (M = 0.67, SD = 0.14), indicating that, on average, the items tend to be moderately easy with a moderate spread in item difficulty. This spread ensures a diverse range of difficulty levels across the items, catering to various proficiency levels of the test-takers.

\begin{table}[htbp]
\centering
\caption{Item difficulty and discrimination indices for each item.}
\label{table-item_selection}
{
\renewcommand{\arraystretch}{1.2}
\begin{tabular}{ccc|ccc}
\hline
\textbf{Item} & \textbf{Difficulty} & \textbf{Discriminative Index} & \textbf{Item} & \textbf{Difficulty} & \textbf{Discriminative Index} \\
\hline
1  & 0.892 & 0.485 & 14 & 0.299 & \textbf{0.272} \\
2  & 0.270 & 0.381 & 15 & 0.363 & \textbf{0.205} \\ 
3  & 0.902 & 0.358 & 16 & 0.696 & 0.489 \\ 
4  & 0.578 & 0.338 & 17 & 0.613 & 0.479 \\ 
5  & 0.613 & 0.432 & 18 & 0.721 & 0.390 \\ 
6  & 0.245 & \textbf{0.064} & 19 & 0.804 & 0.332 \\ 
7  & 0.574 & 0.339 & 20 & 0.721 & \textbf{0.230} \\
8  & 0.853 & 0.448 & 21 & 0.691 & 0.555 \\
9  & 0.304 & \textbf{0.249} & 22 & 0.627 & 0.500 \\ 
10 & 0.676 & 0.354 & 23 & 0.706 & 0.376 \\ 
11 & 0.593 & 0.374 & 24 & 0.799 & 0.378 \\ 
12 & 0.730 & 0.406 & 25 & 0.603 & 0.432 \\ 
13 & 0.485 & 0.355 &    &       &       \\ 
\hline
\end{tabular}
}
\end{table}

\subsubsection{Assumption evaluation}

Both the assumptions of unidimensionality and local independence were confirmed. The single-factor model demonstrated a \(\chi^2\)/\textit{df} ratio of 1.51, which is below the recommended threshold of 2, indicating a good fit for the data. The RMSEA was 0.038, with a 90\% confidence interval ranging from 0.028 to 0.048. This value is well below the criterion of 0.05, suggesting an excellent fit. Additionally, the SRMSR was 0.050, meeting the criterion of less than 0.10. Additionally, the examination of residual correlations revealed that no pairs exceeded the threshold (0.2), indicating that the assumption of local independence was confirmed.

\subsubsection{Structural validity}

For the comparison between the Rasch and the 2PL model, the ANOVA results indicated a significant improvement in fit with the 2PL model \(\chi^2 = 55.596\), \(df = 19\), \(p < 0.001\), suggesting that the added complexity of allowing for varying item discriminations provided a better fit to the data. However, for the comparison between the 2PL and the 3PL model, the ANOVA results showed that the 3PL model did not significantly improve the fit over the 2PL model \(\chi^2 = 23.888\), \(df = 20\), \(p = 0.247\), indicating that accounting for guessing parameters did not significantly enhance the model's explanatory power. 

\begin{table}[ht]
\centering
\caption{Fit indices and information criteria for the Rasch, 2PL, and 3PL models.}
\label{tab-model_fit}
{
\renewcommand{\arraystretch}{1.2}
\begin{tabular}{lccccccc}
\hline
Model & M2 & RMSEA & SRMR & TLI & CFI & AIC & BIC \\
\hline
Rasch & 292.340 & 0.040 & 0.072 & 0.944 & 0.945 & 7823.532 & 7904.246 \\
2PL & 225.325 & 0.031 & 0.052 & 0.967 & 0.970 & 7805.936 & 7959.678 \\
3PL & 166.014 & 0.018 & 0.051 & 0.989 & 0.991 & 7822.048 & 8052.660 \\
\hline
\end{tabular}
}
\end{table}

\begin{figure*}[htbp]
    \centering
    \includegraphics[width=0.99\linewidth]{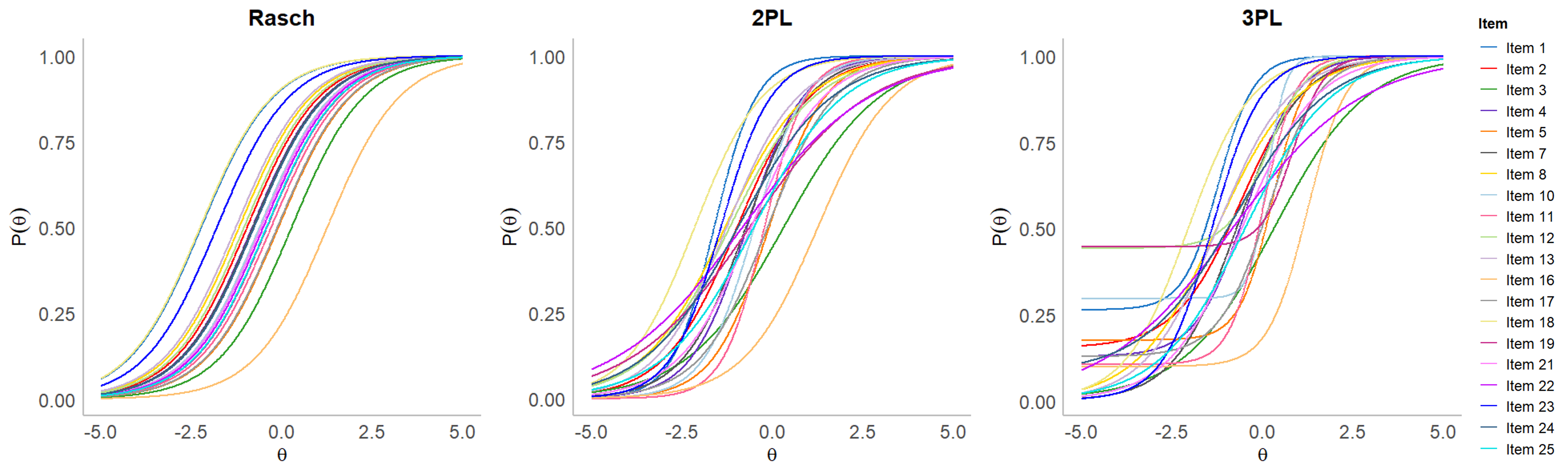}
    \caption{Item characteristic curves for the Rasch, 2PL, and 3PL models. \(\theta\) presents the latent trait, GenAI literacy.}
    \label{fig-icc}
\end{figure*}

As shown in Figure \ref{fig-icc}, the items in all three models demonstrated an S-curve shape, indicating that the probability of a correct response increases with higher levels of the latent trait. These patterns are consistent with expectations for assessments that measure proficiency like GenAI literacy \citep{reise2009item}. Furthermore, as shown in Table \ref{table-item_fit_indices}, the signed chi-squared item fit indices for the Rasch, 2PL, and 3PL models vary across the items. Items 4, 11, and 19 stand out with significant p-values, indicating potential model misfit under the Rasch model. Item 19 also shows a significant misfit under the 2PL model and the 3PL model. In contrast, other items generally exhibit non-significant p-values, suggesting an adequate fit for those items across different models. This item fit evaluation indicates that both 2PL and 3PL models may provide a better fit for the majority of items compared to the Rasch model.

\begin{table}[htbp]
\centering
\caption{Signed chi-squared item fit indices for the Rasch, 2PL, and 3PL models.}
\label{table-item_fit_indices}
{
\renewcommand{\arraystretch}{1.2}
\begin{tabular}{l|ccc|ccc|ccc}
\hline
\textbf{Item} & \multicolumn{3}{c|}{\textbf{Rasch}} & \multicolumn{3}{c|}{\textbf{2-PL}} & \multicolumn{3}{c}{\textbf{3-PL}} \\ \hline
              & $S\text{-}\chi^2$ & df & $p$          & $S\text{-}\chi^2$ & df & $p$         & $S\text{-}\chi^2$ & df & $p$ \\
\hline
1  & 24.164 & 13 &.120 & 18.887 & 11 &.252 & 18.134 & 10 &.212 \\
2  & 15.151 & 10 &.212 & 14.322 & 10 &.393 & 12.708 & 10 &.536 \\
3  & 19.574 & 13 &.198 & 16.637 & 12 &.393 & 15.409 & 11 &.471 \\
4  & 29.801 & 13 & \textbf{.033} & 19.875 & 15 &.393 & 16.147 & 13 &.536 \\
5  & 21.876 & 13 &.163 & 21.642 & 13 &.252 & 22.385 & 12 &.212 \\
7  & 24.593 & 13 &.120 & 14.831 & 15 &.639 & 13.813 & 14 &.714 \\
8  & 8.871  & 13 &.824 & 5.632  & 12 &.974 & 6.141  & 11 &.909 \\
10 & 7.183  & 13 &.892 & 5.671  & 14 &.974 & 6.600    & 13 &.922 \\
11 & 30.991 & 13 & \textbf{.030} & 25.162 & 13 &.220 & 27.214 & 13 &.120 \\
12 & 15.094 & 14 &.499 & 14.616 & 14 &.623 & 14.661 & 13 &.585 \\
13 & 18.716 & 12 &.198 & 16.549 & 14 &.518 & 15.418 & 11 &.471 \\
16 & 10.928 & 13 &.686 & 10.023 & 13 &.807 & 9.903  & 12 &.805 \\
17 & 13.555 & 11 &.398 & 13.135 & 11 &.518 & 7.847  & 11 &.808 \\
18 & 13.998 & 13 &.499 & 14.315 & 13 &.587 & 14.259 & 12 &.568 \\
19 & 43.512 & 14 & \textbf{<.001} & 39.537 & 14 & \textbf{<.001} & 32.063 & 13 & \textbf{.040} \\
21 & 23.203 & 13 &.130 & 21.786 & 12 &.252 & 12.172 & 11 &.585 \\
22 & 20.463 & 13 &.198 & 10.580  & 11 &.639 & 8.435  & 10 &.805 \\
23 & 13.160 & 14 &.642 & 10.484 & 14 &.807 & 10.126 & 13 &.805 \\
24 & 12.111 & 14 &.686 & 12.717 & 14 &.686 & 10.509 & 13 &.805 \\
25 & 16.962 & 11 &.198 & 17.441 & 11 &.317 & 19.571 & 11 &.212 \\
\hline
\end{tabular}
}
\end{table}

Based on these analyses, the 2PL model was determined to be the best-fitting model. The fit indices and information criteria for the 2PL model indicated a robust structural validity (Table \ref{tab-model_fit}): the M2 statistic was 225.325, and the RMSEA was 0.031, indicating a good fit. The SRMSR was 0.052, which also suggests a good fit, while both the TLI and CFI were 0.967 and 0.970, respectively, suggesting an excellent fit. Additionally, the AIC and BIC were 7805.936 and 7959.678, respectively, supporting the 2PL model as more favourable when balancing fit and model complexity.

\subsubsection{Reliability}

\begin{figure*}[htbp]
    \centering
    \includegraphics[width=0.65\linewidth]{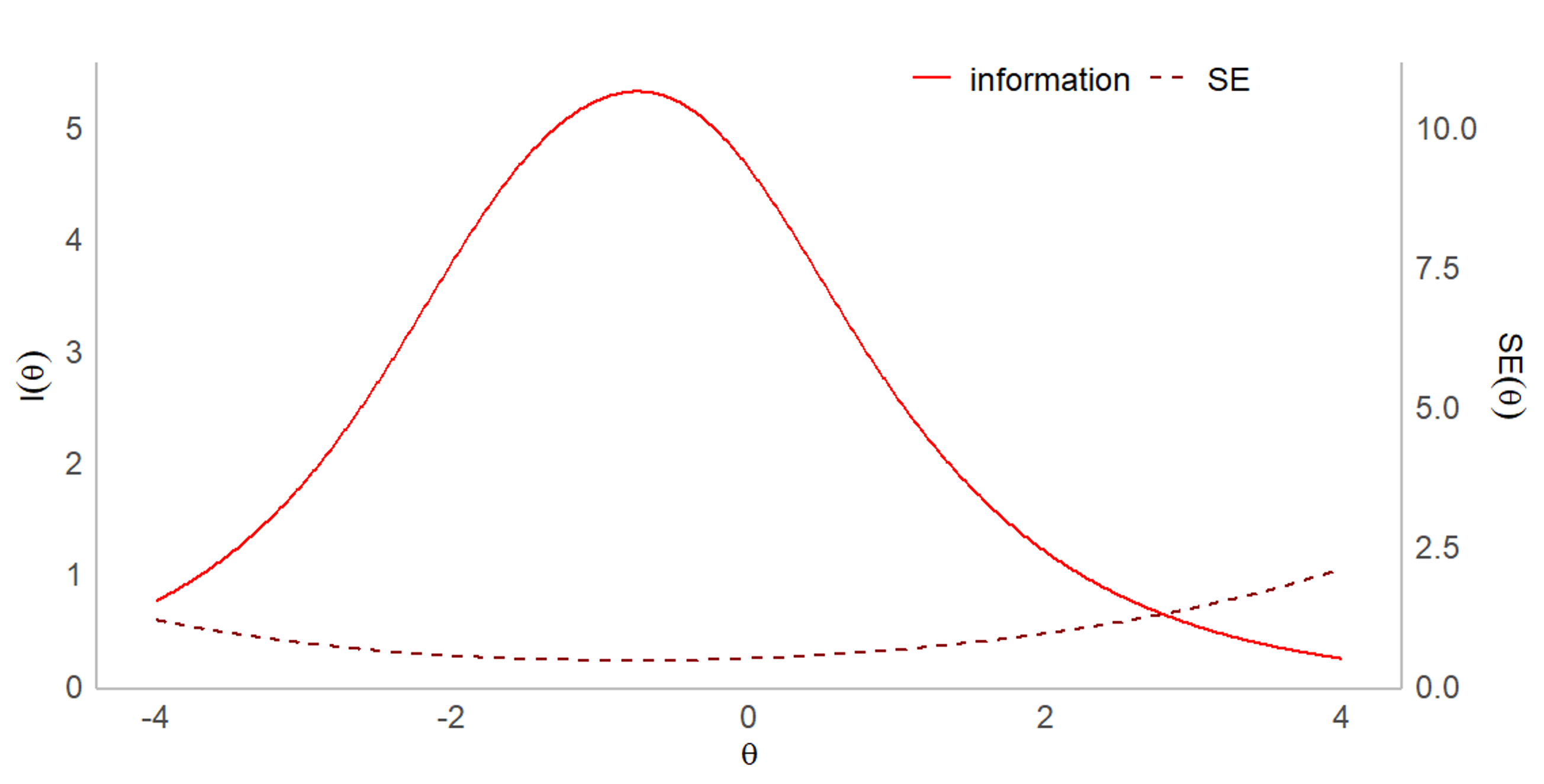}
    \caption{Test information function for the 2PL models. \(\theta\) presents the latent trait, GenAI literacy.}
    \label{fig-tif}
\end{figure*}

The reliability of the GLAT was assessed by examining internal consistency using Cronbach’s alpha and omega total. Cronbach’s alpha was calculated to be 0.80, indicating good reliability, as it surpassed the threshold of 0.7 commonly used in educational testing \citep{miller1995coefficient}. In addition, the coefficient omega total was computed to further evaluate reliability, resulting in a value of 0.81. This suggested a high level of internal consistency and confirmed the reliability of the GLAT in measuring GenAI literacy, particularly given the varying loadings on the construct \citep{dunn2014alpha}. Additionally, the precision and reliability of the GLAT were further examined using the test information function. As Figure \ref{fig-tif} shows, the GLAT provided the most information at a proficiency level of \( \theta = -0.8 \), indicating that the test is highly reliable for individuals with below-average GenAI literacy. The maximum information value of 5.31 at this point suggests strong measurement precision for this target group. However, the information decreases for proficiency levels further from \( \theta = -0.8 \), especially for individuals with higher proficiency, where the test is less discriminative. The standard error (SE) was also lowest around \( \theta = -0.8 \), further supporting the test’s high precision for low to moderate GenAI literacy, while precision diminished for more extreme proficiency levels. 

\subsection{External Validity (RQ2)}

The predictive model was statistically significant, \( F(4, 78) = 9.233 \), \( p < .001 \), with an \( R^2 \) of 0.321, indicating that the model accounted for approximately 32.1\% of the variance in \textit{AI-assisted score}. Among the predictors, \textit{GLAT-literacy} (\( \beta = 0.220 \), \( t = 2.093 \), \( p = .040 \)) and \textit{VLAT-literacy} (\( \beta = 0.322 \), \( t = 2.946 \), \( p = .004 \)) were both significant positive predictors of \textit{AI-assisted score}. For a standard deviation increase in \textit{GLAT-literacy}, the \textit{AI-assisted score} increased by approximately 0.220 standard deviations, suggesting that greater proficiency in GenAI literacy is associated with enhanced performance in tasks supported by the GenAI chatbot. Similarly, a standard deviation increase in \textit{VLAT-literacy} resulted in an increase of approximately 0.322 standard deviations in the \textit{AI-assisted score}, indicating that domain knowledge, such as visualisation literacy, significantly contributes to improved task performance, which is expected considering the task involved comprehending visual analytics. Whereas, \textit{ChatGPT-literacy} showed a negative relationship with the \textit{AI-assisted score} (\( \beta = -0.159 \)), but was not significant \( t = -1.579 \), \( p = 0.118 \)). This suggests students' self-reported proficiency with ChatGPT was not a significant factor in predicting their performance scores in GenAI-assisted tasks. The baseline score (\( \beta = 0.098 \), \( t = 0.907 \), \( p = 0.367 \)) did not significantly predict the \textit{AI-assisted score}, suggesting that initial task performance without AI assistance did not substantially influence outcomes when using the GenAI chatbot. This highlights the independent contributions of \textit{GLAT-literacy} and \textit{VLAT-literacy} to learners' task performance.
\section{Discussion}

Effective and valid measurement of GenAI literacy is essential in higher education as learners and educators increasingly encounter GenAI tools in their study, work, and daily lives \citep{yan2024promises, cukurova2024interplay, khosravi2023generative}. This study developed and validated the GLAT in line with established standards for psychological and educational measurement \citep{thorndike1991measurement, aera2014standards}. Regarding RQ1, the GLAT demonstrated a 2PL model with strong structural validity, meeting the requirements of item discrimination and difficulty across a diverse sample. This indicates that the GLAT effectively differentiates individuals with varying GenAI literacy levels, which is crucial for accurately assessing competencies related to GenAI use \citep{annapureddy2024generative, yan2024promises, zhao2024chatgpt, Bozkurt_2024}. The GLAT also showed good reliability, particularly in assessing students with low to moderate GenAI literacy. This aligns with its intended use, considering the current state of GenAI literacy, where the technology is relatively new and integrated training is limited in higher education curricula \citep{holmes2023guidance, jin2024generative}. The GLAT is thus especially valuable for identifying individuals who may need additional education and support to effectively understand and use GenAI ethically. These findings highlight the GLAT's utility in assessing foundational GenAI knowledge, particularly where students have limited prior exposure or training. However, as GenAI training becomes more integrated into higher education and students' average GenAI literacy improves, the GLAT will require updates to remain relevant. This aligns with the need for an iterative design process for test instruments to adapt to new data and evolving use contexts \citep{aera2014standards}.

In terms of RQ2 and external validity, we examined the extent to which the GLAT predicts learners' performance in tasks involving GenAI chatbots compared to self-reported instruments, using visualisation literacy and baseline performance as control variables. The predictive model showed that GenAI literacy, as measured by GLAT, was a significant predictor of learners' performance in GenAI-supported tasks, whereas domain knowledge (e.g., visualisation literacy) served as a control to account for differences in learners' comprehension of visual information. The significant positive relationship between GLAT and AI-assisted task performance underscores the value of reliably assessing GenAI literacy to predict real-world learner outcomes \citep{annapureddy2024generative, chiu2024future}. In contrast, self-reported ChatGPT proficiency was not a significant predictor, highlighting the limitations of self-assessment, which may be prone to biases or inaccuracies \citep{Lintner_2024, ng2021conceptualizing}. The control for domain knowledge (e.g., visualisation literacy) ensured that the observed effects were specific to GenAI literacy, thereby reinforcing the importance of targeted skill development in GenAI \citep{lee2024development, lyu2024evaluating}. These findings suggest that enhancing GenAI literacy has a direct effect on learners' ability to effectively engage with GenAI tools, independent of their domain knowledge. This insight is critical for educators aiming to design targeted interventions that bolster students' competencies in using GenAI technologies effectively in diverse educational contexts.

\subsection{Implications to Research and Practice}
 
The study's findings have profound implications for advancing research and practice in GenAI literacy within higher education. The development of the GLAT underscores the need for performance-based measures over traditional self-reported assessments, addressing the limitations and biases inherent in self-assessment tools \citep{ng2021conceptualizing, Lintner_2024}. This shift to more reliable measures will enable educators and researchers to make informed decisions about integrating GenAI into educational settings. For educators, leveraging the GLAT offers a diagnostic tool to assess and enhance students' GenAI literacy, highlighting individual needs for targeted interventions \citep{alzubi2024generative, bozkurt2023unleashing}. The assessment's external validity in GenAI-supported learning tasks further underscores its practical utility, providing insights into students' preparedness to navigate GenAI technologies in diverse educational contexts. By incorporating the GLAT into curriculum development, educators can better align teaching strategies with the specific competencies required for effective GenAI engagement, thereby preparing students for an AI-driven future. Furthermore, the study encourages researchers to adopt a comprehensive, iterative approach to developing and validating educational assessments, ensuring their continued relevance amidst the evolving GenAI landscape \citep{holmes2023guidance, jin2024generative}. By focusing on multidimensional literacy frameworks that integrate foundational knowledge, practical skills, and ethical understanding, future research can enhance the robustness of educational instruments and cultivate a nuanced understanding of GenAI literacy in academia \citep{zhao2024chatgpt, Bozkurt_2024}.

\subsection{Limitations and Future Directions}

While this study advances the measurement of GenAI literacy in higher education, several limitations warrant attention. Firstly, the GLAT was developed and validated primarily with higher education students, excluding younger K-12 students and educators. Future research should extend its applicability across different educational levels and include educators to ensure broader relevance and utility. Additionally, the study's examination of external validity is based on context-specific tasks involving visual analytics and chatbot interactions. Future investigations should incorporate various contexts and task complexities to comprehensively understand how GenAI literacy affects learning performance across different domains. A notable limitation is the focus on certain types of domain knowledge, such as visualisation literacy, without considering other relevant knowledge areas that may affect task performance. Thus, future studies should examine a wider range of domain knowledge to better control its influence on GenAI literacy outcomes. The rapidly evolving nature of GenAI technology presents another limitation. As these tools advance, so too must the instruments assessing GenAI literacy. Researchers should continuously update and refine the GLAT to keep pace with new developments and maintain its effectiveness and accuracy \citep{aera2014standards}. Lastly, it is important to note that the test is conducted in English, which may limit its accessibility and relevance for non-English-speaking participants. Future studies should explore the adaptation of the assessment for different languages to ensure its validity and applicability across diverse linguistic populations.
\section{Conclusion}

This study introduced the GLAT, a performance-based instrument designed to assess GenAI literacy within higher education contexts. The GLAT demonstrated robust structural validity and reliability, particularly in evaluating foundational GenAI knowledge among students with varying levels of expertise. The external validity of the GLAT further underscored its practical utility, showcasing a significant positive relationship between GLAT scores and learners' performance in GenAI-supported tasks. This study advocates for the integration of performance-based assessments in addition to traditional self-reported measures to evaluate GenAI literacy reliably. The findings highlight the need for continuous adaptation of assessment tools to keep pace with technological advancements, thereby equipping educators and students with the skills necessary to engage in an AI-driven future effectively. Future research should focus on expanding the applicability of the GLAT across diverse educational levels and contexts, addressing the complex and evolving landscape of GenAI technologies.

\bibliographystyle{cas-model2-names}

\bibliography{0_reference}



\end{document}